\documentstyle[preprint,aps,prb]{revtex}
\begin{document}
\draft
\title{ Charge dynamics and optical conductivity
of the $t-J$ model}
\author{G. Jackeli and N. M. Plakida}
\address{Joint Institute for Nuclear Research, Dubna 141980, Russia }
\date{\today}
\maketitle
\begin{abstract}
\widetext
The dynamic charge susceptibility and the optical conductivity are 
calculated in the planar $t-J$ model within the memory function method, 
working directly
in terms of Hubbard operators.
The  density fluctuation  spectrum consists of a damped sound-like 
mode for small wave vectors  and a broad high energy peak ($\sim t$)
for large momenta.
The study of the optical conductivity  shows that 
electron scattering from spin fluctuations
leads to the Drude-frequency dependent relaxation rate 
which exhibits a crossover from $\omega^{3/2}$ behavior at low
frequencies ($\omega <2|\mu|$), to a linear $\omega$--dependence for
frequencies larger than $2|\mu|$.
Due to the spin-polaron nature of charge carriers, extra absorptions
arise starting at a frequency $\omega \agt J$.
The obtained results are in a good agreement with 
exact diagonalization studies.
\end{abstract}

\narrowtext

\newpage
\section{Introduction}

Among  other  unconventional normal state
properties of high-$T_{\text{c}}$ superconductors,
an anomalous charge dynamics has also been detected 
in the optical measurements of the  underdoped samples. \cite{brening}
Namely, a non-Drude fall-off of the low-frequency absorption indicating
a linear $\omega$--dependence of the relaxation rate 
and an anomalous  mid-infrared (MIR) band with a typical energy
$\sim 0.1$ eV have been observed. \cite{thomas,uchida} 

It is widely believed that unusual  properties of the
superconducting cuprates are  due to the
strong electron correlations. \cite{brening}
The minimal model to describe correlation effects in the cuprates is 
the $t-J$ model. 
While a number of analytic works have been carried out to
investigate spin dynamics within the $t-J$ model,
only  few of authors have
studied charge dynamics. \cite{wang,zeyher,horsch}
In Refs.~\onlinecite{wang,zeyher} charge fluctuations have been studied by
the slave boson and Hubbard operator (HO) formalism within the
leading order of $1/N$ expansion, respectively.
It was found that the density fluctuations
at large momenta show a sharp high-energy
peak corresponding to the collective mode which reduces
to the sound mode in the long-wavelength limit. \cite{wang,zeyher}
Later, the authors of Ref.~\onlinecite{horsch}
showed that  next-order corrections in the $1/N$ expansion
lead to the broadening of the high-energy peak due to incoherent
motion of bear holes. 
Similar features of the density response have been previously observed
in  exact diagonalization studies of small clusters. \cite{tohoyama}
Recently, the charge correlations in the AFM phase of the
$t-J$ model has been investigated by commulant version of projection
technique \cite{becker}.

In the present paper, we investigate the charge fluctuation spectrum of
the $t-J$ model in the paramagnetic state 
with short-range antiferromagnetic (AFM) correlations.
We develop a self-consistent theory for the dynamic charge
susceptibility (DCS) by applying 
the memory function method in terms of HO's.
The employment of the HO technique has a twofold advantage;
By using the equations of motion for the HO's we
automatically take into account scattering of electrons on spin and charge
fluctuations originated from the strong correlations, as it has first
been pointed out by Hubbard. \cite{hubbard} Moreover, the HO formalism
allows us to preserve rigorously the local constraint
of no double occupancy.

We calculate the memory function within the mode coupling approximation
(MCA)
in terms of the dressed particle-hole (p-h) and spin fluctuations.
Similarly to the nearly antiferromagnetic Fermi liquid approach, \cite{pines1}
we treat fermionic and localized spin excitations as
independent degrees of freedom. 
We show that the memory function 
involves two contributions. The first one stems from the hopping term
and describes a particle-hole contribution from the itinerant hole
subsystem. The second one involves scattering processes 
of electrons on charge and spin fluctuations and comes both from
kinematics and exchange interactions. 

Further, we perform an analytic analysis of  different limiting 
behavior of DCS to show that the essential features observed in the
exact diagonalization studies \cite{tohoyama} can be reproduced within the present formalism.
We find out that for small $q$ the DCS is mainly governed by the sound
 mode.
Although the unrenormalized sound velocity is larger than  
 the Fermi velocity, unlike the Fermi liquid theory,
 the ``self-energy'' corrections
 lead to 
softening of the sound. 
The renormalized sound falls down into the p-h continuum  getting 
a finite damping due to the decay into pair excitations.
In the short-wavelength limit, 
density fluctuation spectrum mainly consists of a 
broad high-energy peak.
At large enough wave vectors the peak is dispersed out of 
the  coherent p-h continuum  and broadens 
due to  high energy ($\sim t$)  transitions
 involving the incoherent band  of one-particle excitations.
 
We also discuss the optical conductivity $\sigma(\omega)$. 
For low frequencies we analyze $\sigma(\omega)$ in terms of the
generalized Drude 
law. We show that there is  a mass enhancement of order
$m^{*}/m\simeq 6$, due to the
electron scattering on spin fluctuations. These scattering
processes also lead to a frequency-dependent relaxation rate 
which exhibits a crossover from $\omega^{3/2}$ behavior at low
frequencies, $\omega <2|\mu|$, to a linear $\omega$--dependence for
$\omega>2|\mu|$. A possible origin of 
the MIR band is also discussed.

The paper is organized as follows. In the next section, 
we give the basic definitions and sketch the memory function formalism. In Sec. III, we employ  MCA
to calculate a memory function. The dynamic charge susceptibility and 
optical conductivity are discussed in Secs. IV and V, respectively.
The last section summerizes our main results.

\section{ Model and memory function formalism}
The $t-J$ model expressed in terms of HO's, 
$X_{i}^{\alpha\beta}=|i,\alpha\rangle\langle i,\beta|$,
 reads as
\begin{eqnarray}
H&=&H_{t}+H_{J}=
-\sum_{ i,j}t_{ij} X_{i}^{\sigma 0}X_{j}^{0\sigma}\nonumber\\
&+&\frac{1}{4} \sum_{ i,j,\sigma}J_{ij} \{X_{i}^{\sigma\bar{\sigma}}
X_{j}^{\bar{\sigma}\sigma}-X_{i}^{\sigma\sigma}
X_{j}^{\bar{\sigma}\bar{\sigma}}\},
\label{1}
\end{eqnarray}
where the indices $0$ and $\sigma=\pm 1$ correspond to a hole
and an electron with  spin $\sigma/2$, respectively,
$t_{ij}=t$ and $J_{ij} =J$ for the nearest--neighbor (n.n.) sites on a
planar lattice.
The HO's can be either Bose-like or Fermi-like and obey the following
 on-site multiplication  rules
$
X_{i}^{\alpha\beta}X_{i}^{\gamma\delta}=\delta_{\beta\gamma}
X_{i}^{\alpha\delta}
$
and the commutation relations
\begin{equation}
\left[X_{i}^{\alpha\beta}X_{j}^{\gamma\delta}\right]_{\pm}=
\delta_{ij}\left(\delta_{\beta\gamma}X_{i}^{\alpha\delta}\pm
\delta_{\delta\alpha}X_{i}^{\gamma\beta}\right),
\label{2}
\end{equation}
where the upper sign stands for the case when both
HO's are Fermi-like,  otherwise the lower sign
should be adopted. In the $t-J$ model only singly occupied sites are
retained and the completeness relation for the HO's reads as
\begin{equation}
X_{i}^{00}+\sum_{\sigma}X_{i}^{\sigma\sigma}=1.
\label{3}
\end{equation}
The spin and density operators are expressed by
HO's as
\begin{equation}
S_{i}^{\sigma}=X_{i}^{\sigma\bar{\sigma}},\;\;\;
S_{i}^{z}=\frac{1}{2}\sum_{\sigma}\sigma X_{i}^{\sigma\sigma},\;\;\;
n_{i}=\sum_{\sigma}X_{i}^{\sigma\sigma}.
\label{4}
\end{equation}

The  dynamic charge susceptibility
$N_{\bf q}(\omega)$ is given by  a 
Fourier transformed  two-time retarded Green function (GF) \cite{zubarev}
\begin{equation}
N_{\bf q}(\omega)=-\langle\langle
n_{\bf q}|n_{-{\bf q}}\rangle\rangle_{\omega}=i
 \int_{0}^{\infty}dt e^{i\omega t}
\langle[n_{\bf q}(t),n_{-{\bf q}} ]\rangle.
\label{5}
\end{equation}

To calculate $N_{\bf q}(\omega)$, we  employ the memory formalism
as discussed in Refs.~\onlinecite{forster,gotze,plakida1}.
First we introduce  density-density relaxation function 
\begin{equation}
\Phi_{q}(\omega)=((n_{\bf q}|n_{-{\bf q}} ))_{\omega} = - i \int_{0}^{\infty} 
dt e^{i\omega t}(n_{\bf q}(t)|n_{-{\bf q}} ),
\label{6}
\end{equation}
where the  Kubo-Mori scalar product is defined as
\begin{equation}
(A(t), B) =\int_{0}^{\beta}d\lambda \langle A(t-i\lambda) B \rangle,
\label{7}
\end{equation}
with $\beta=1/T$.
The DCS $N_{\bf q}(\omega)$  is coupled to the relaxation function 
$\Phi_{q}(\omega)$ 
by the 
equation 
\begin{equation}
N_{\bf q}(\omega)=N_{\bf q}-\omega\Phi_{q}(\omega),
\label{8}
\end{equation}
where $N_{\bf q}=N_{\bf q}(0)$ is the static susceptibility.

We  introduce  the memory function $M_{\bf q}^{0}(\omega)$ for the 
relaxation function $\Phi_{\bf q}(\omega)$  as
\begin{equation}
\Phi_{\bf q}(\omega)=\frac{N_{\bf q}}
{\omega - M_{\bf q}^{0}(\omega)/N_{\bf q}}.
\label{9}
\end{equation} 
By adopting the equation of motion method for the relaxation function 
$\Phi_{\bf q}(\omega)$
one finds that the memory function $M_{\bf q}^{0}(\omega)$
is given by the irreducible
part of the relaxation function for ``currents'' \cite{tser,irr}
\begin{equation}
M_{\bf q}^{0}(\omega)=((j_{\bf q}|j_{-{\bf q}}))_{\omega}^{irr}.
\label{10}
\end{equation}
The ``current'' operator $j_{\bf q}$ in the
site representation reads as
\begin{equation}
j_{i}={\dot n_{i}}=-i[n_{i},H]=-i\sum_{j,\sigma}t_{ij}(X_{j}^{\sigma 0}X_{i}^{0\sigma}-\mbox{H.c.}).
\label{11}
\end{equation}
The Heisenberg part of the Hamiltonian (\ref{1}) 
conserves the local particle number
and thus gives no contribution to the ``current'' operator (\ref{11}).
To treat properly  a contribution from the  $H_{J}$ term, we 
go one step further and
similarly to Eq.(\ref{9}) we introduce the  memory function $M_{\bf q}(\omega)$
for the relaxation function for ``currents''
 $M_{\bf q}^{0}(\omega)$ (\ref{10}), by the following equation
\begin{equation}
M_{\bf q}^{0}(\omega)=\frac{m_{\bf q}}
{\omega - M_{\bf q}(\omega)/m_{\bf q}},
\label{12}
\end{equation}
where  
\begin{equation}
m_{\bf q}=-\langle\langle
j_{\bf q}|j_{-{\bf q}}\rangle\rangle_{\omega=0}=i\langle
[j_{\bf q},n_{-{\bf q}}]\rangle 
\label{15}
\end{equation}
is the first moment of DCS and 
the memory function $M_{\bf q}(\omega)$ 
is given by the irreducible part of 
the relaxation function for ``forces '':
\begin{equation}
M_{\bf q}(\omega)=((F_{\bf q}|F_{-{\bf q}}))_{\omega}^{irr}
\label{13}
\end{equation}
with 
\begin{equation}
F_{\bf q}=\dot{j}_{\bf q}=-i[j_{\bf q},H].
\label{14}
\end{equation}

 Further, to close the system of equations
 [see Eqs.(\ref{8}),(\ref{9}), and (\ref{12})],
 we employ a mode coupling
approximation for the memory function 
$M_{\bf q}(\omega)$.

\section{Mode coupling approximation}
First we express the memory function in terms of the irreducible part of 
time-dependent corelation function for ``forces''
by means of the fluctuation-dissipation theorem \cite{forster}
\begin{eqnarray}
M_{\bf q}(\omega)&=&
\int\limits_{-\infty }^{\infty }
\frac{d\omega^{\prime}}{2\pi}
\frac{e^{\beta\omega^{\prime}}-1}{\omega^{\prime}(\omega-\omega^{\prime}+i\eta)}\nonumber\\
&\times&\int\limits_{-\infty }^{\infty }dte^{-i\omega^{\prime}t}
\langle F_{-{\bf q}}(t)|F_{{\bf q}}\rangle^{irr}.
\label{16}
\end{eqnarray}
The ``force''operator is given by
\begin{equation}
F_{i}=\sum_{m,j,\sigma\sigma^{\prime}}\left[\Pi_{i,m,j}^{\sigma\sigma^{\prime}}-
\Pi_{m,i,j}^{\sigma\sigma^{\prime}}+\mbox{H.c.}\right],
\label{17}
\end{equation}
\begin{eqnarray}
\Pi_{m,i,j}^{\sigma\sigma^{\prime}}=
 t_{im}&[&t_{mj}(X^{\sigma 0}_{i} X^{ 0 \sigma }_{j}
\delta_{\sigma\sigma^{\prime}}
-X^{\sigma 0}_{i} X^{ 0 \sigma^{\prime}}_{j}
 B_{m}^{\sigma^{\prime}\sigma})\nonumber\\
&+&J_{mj}X^{\sigma 0}_{i} X^{ 0 \sigma^{\prime}}_{m}
 B_{j}^{\sigma^{\prime}\sigma}],
\label{18}
\end{eqnarray}
where the Bose-like operator
\begin{eqnarray}
B_{i}^{\sigma^{\prime}\sigma}&=&X^{\bar{\sigma}\bar{\sigma}}_{i}
\delta_{\sigma\sigma^{\prime}}-
X^{\bar{\sigma}\sigma}_{i}
\delta_{\bar{\sigma}\sigma^{\prime}}\nonumber\\
&=&\left[\frac{1}{2}n_{i}-\sigma S_{i}^{z}\right]
\delta_{\sigma\sigma^{\prime}}
-S_{i}^{\bar{\sigma}}
\delta_{\bar{\sigma}\sigma^{\prime}}
\label{19}
\end{eqnarray}
describes electron scattering on spin and charge fluctuations.

The sum in Eq.(\ref{17}) contains the 
products of HO's from the same site.
As follows, such products give no contribution 
to the memory function, while being decoupled 
they do contribute. That is a result of the complexity of HO's algebra
(\ref{2}).
To show this, let us consider the  term 
given by Eq.(\ref{18});
since ($i$,$m$) and ($m$,$j$) are  n.n. pairs $i\neq m$ and
 $m\neq j$, however $i$ can be equal to $j$.
For the latter case, $i=j$, the first term in Eq.(\ref{18}) 
is linear in  the density operator
$(X^{\sigma 0}_{i} X^{ 0 \sigma }_{i}=X^{\sigma\sigma}_{i})$ 
and thus gives no contribution to the irreducible part 
of the correlation function for ``forces''. \cite{irr}
As for the second term (\ref{18}), 
one can easily verify that in the case  $i=j$ it is canceled out
by its counter part from the sum (\ref{17}).
Finally, the last term in Eq.(\ref{18}) vanishes for $i=j$
since $ X^{\sigma 0}_{i} B_{i}^{\sigma^{\prime}\sigma}=0$
 due to the constraint.
Therefore we have to subtract these terms from the ``force'' operator.
As a result, we 
come to the following expression in the momentum space 
\begin{eqnarray}
F_{\bf q}&=&-\frac{1}{\sqrt{N}}\sum_{{\bf k},\sigma}G_{{\bf k},{\bf q}}
X^{\sigma 0}_{{\bf k}+{\bf q}} X^{ 0 \sigma }_{\bf k}\nonumber\\
&-&\frac{1}{N}\sum_{{\bf k},{\bf p},\sigma\sigma^{\prime}}M_{{\bf k},{\bf q},{\bf p}}
X^{\sigma 0}_{{\bf k}+{\bf q}-{\bf p}} X^{ 0 \sigma^{\prime}}_{\bf k}
 B_{\bf p}^{\sigma^{\prime}\sigma},
\label{20}
\end{eqnarray}
the vertices $G_{{\bf k},{\bf q}}$ and $M_{{\bf k},{\bf q},{\bf p}}$ 
are given by
\begin{eqnarray}
G_{{\bf k},{\bf q}}&=&g_{{\bf k},{\bf q}}-\overline{g}_{{\bf k},{\bf q}}
\label{21}\\
M_{{\bf k},{\bf q},{\bf p}}&=&\sum_{i}[m_{{\bf k},{\bf q},{\bf p}}^{(i)}-
\overline{m}_{{\bf k},{\bf q},{\bf p}}^{(i)}]
\label{22}
\end{eqnarray}
where
\begin{eqnarray}
g_{{\bf k},{\bf q}}&=&(zt)^{2}\gamma_{{\bf k},{\bf q}}^{2},\;\;
m_{{\bf k},{\bf q},{\bf p}}^{(1)}=\frac{z^{2}tJ}{2}\gamma_{\bf p}
[\gamma_{{\bf k}-{\bf p},{\bf q}}-\gamma_{{\bf k},{\bf q}}]\nonumber\\
& & m_{{\bf k},{\bf q},{\bf p}}^{(2)}=(zt)^{2}
[\gamma_{{\bf k},{\bf q}}\gamma_{{\bf k}+{\bf q}-{\bf p}}
-\gamma_{{\bf k}-{\bf p},{\bf q}}\gamma_{\bf k}].
\label{23}
\end{eqnarray}
In Eqs.(\ref{21}) and (\ref{22}) $\overline{g}$($\overline{m}$)  
denote $g$($m$) averaged over the Brillouin zone and are
given by
\begin{eqnarray}
\overline{g}_{{\bf k},{\bf q}}&=&2zt^{2}(1-\gamma_{\bf q}),\;
\overline{m}_{{\bf k},{\bf q},{\bf p}}^{(1)}=\frac{ztJ}{2}
[\gamma_{{\bf k},{\bf q}}-\gamma_{{\bf k}-{\bf p},{\bf q}}]
\nonumber\\
& & \overline{m}_{{\bf k},{\bf q},{\bf p}}^{(2)}=
2zt^{2}\gamma_{{\bf p}-{\bf q},{\bf q}},
\label{24}
\end{eqnarray}
where $\gamma_{{\bf k},{\bf q}}=\gamma_{{\bf k}+{\bf q}}-\gamma_{\bf k}$,
$\gamma_{\bf q}=1/2[\cos(q_{x})+\cos(q_{y})]$ 
and $z=4$ for a 2-dimensional square lattice. 
This  form of the renormalized vertices (\ref{21})-(\ref{22}) 
insures that all the operators in the products of Eq.(\ref{20}) are
from different sites. Therefore HOs
can be simply permuted within the decoupling procedure.

To calculate the irreducible part of the 
time-dependent corelation function in the right-hand side of
Eq.(\ref{16}), we apply the mode-coupling approximation \cite{gotze} in terms 
of an independent propagation of dressed p-h pairs and charge-spin fluctuations.
The proposed approximation is defined by the 
following decoupling of the time-dependent 
correlation functions
\begin{eqnarray}
\langle X_{{\bf k}-{\bf q}}^{\sigma 0}(t)X_{\bf k}^{0\sigma}(t)|
X_{{\bf k}^{\prime}+{\bf q}}^{\sigma^{\prime} 0}
X_{{\bf k}^{\prime}}^{0\sigma^{\prime}}\rangle
\simeq\nonumber\\
\delta_{\sigma,\sigma^{\prime}}\delta_{{\bf k}-{\bf q},{\bf k}^{\prime}}
\langle X_{{\bf k}-{\bf q}}^{\sigma 0}(t)
X_{{\bf k}^{\prime}}^{0\sigma^{\prime}} \rangle
\langle X_{\bf k}^{0\sigma}(t)X_{{\bf k}^{\prime}+
{\bf q}}^{\sigma^{\prime} 0} \rangle,
\label{25} 
\end{eqnarray}
\begin{eqnarray}
\langle X^{\sigma 0}_{{\bf k}-{\bf q}-{\bf p}}(t)
X^{ 0 \sigma^{\prime}}_{\bf k}(t)
 B_{\bf p}^{\sigma^{\prime}\sigma}(t)|
X^{s 0}_{{\bf k}^{\prime}+{\bf q}-{\bf p}^{\prime}}X^{ 0 s^{\prime}}_{{\bf k}^{\prime}}
 B_{{\bf p}^{\prime}}^{s^{\prime}s}
\rangle\nonumber\\
\simeq \delta_{\sigma,s^{\prime}}\delta_{\sigma^{\prime},s}
\delta_{{\bf k}-{\bf q}-{\bf p},{\bf k}^{\prime}}
\delta_{{\bf p},-{\bf p}^{\prime}}\nonumber\\
 \langle X^{\sigma 0}_{{\bf k}-{\bf q}-{\bf p}}(t)
X^{ 0 s^{\prime}}_{{\bf k}^{\prime}} \rangle
 \langle X^{ 0 \sigma^{\prime}}_{\bf k}(t)X^{s 0}_{{\bf k}^{\prime}
+{\bf q}-{\bf p}^{\prime}} \rangle
 \langle B_{\bf p}^{\sigma^{\prime}\sigma}(t)B_{{\bf p}^{\prime}}^{s^{\prime}s} \rangle.
\label{26}
\end{eqnarray}
By using the decoupling scheme and the spectral representation for the
two-time retarded GF's \cite{zubarev} we obtain for the memory function 
\begin{eqnarray}
M_{\bf q}(\omega)=\frac{1}{\omega}
\left[\Pi({\bf q},\omega)-\Pi({\bf q},0)\right],\nonumber\\
\Pi({\bf q},\omega)=\Pi_{1}({\bf q},\omega)+\Pi_{2}({\bf q},\omega),
\label{27}
\end{eqnarray}
where $\Pi_{1}({\bf q},\omega)$ and $\Pi_{2}({\bf q},\omega)$
stem from the first and the second term of Eq.(\ref{20}), 
respectively. Their imaginary
parts are given by
\begin{eqnarray}
\Pi_{1}^{\prime\prime}({\bf q},\omega)&=&\frac{-2\pi}{ N}
\sum_{{\bf k}}G^{2}_{{\bf k},{\bf q}}
\int\limits_{-\infty }^{\infty }
d\omega_{1}n_{\omega_{1},\omega}\nonumber\\
&\times& A_{{\bf k}}(\omega_{1})A_{{\bf k}+{\bf q}}(\omega_{1}+\omega),
\label{28}
\end{eqnarray}
\begin{eqnarray}
\Pi_{2}^{\prime\prime}({\bf q},\omega)&=&\frac{-2\pi}{ N^{2}}
\sum_{{\bf k},{\bf p}}M^{2}_{{\bf k},{\bf q},{\bf p}}
\int\!\!\!\!\int\limits_{-\infty }^{\infty }
d\omega_{1}d\omega_{2}N_{\omega,\omega_{1},\omega_{2}}\nonumber\\
&\times& A_{{\bf k}+{\bf q}-{\bf p}}(\omega-\omega_{1}+\omega_{2})
A_{{\bf k}}(\omega_{2})
\chi_{cs}^{\prime\prime}
( { \bf p},\omega_{1}),
\label{29}
\end{eqnarray}
where $n_{\omega_{1},\omega}=n(\omega_{1})-n(\omega_{1}+\omega)$,
$N_{\omega,\omega_{1},\omega_{2}}=[1+N(\omega_{1})+N(\omega-\omega_{1})]
n_{\omega_{2},\omega-\omega_{1}}$ with $ n(\omega)$ and $N(\omega)$
being Fermi and Bose distribution functions, respectively, and
\begin{equation}
  A_{\bf k}(\omega) =  - {\frac{1}{\pi}}
   \mbox{Im} \langle\!\langle X^{0\sigma}_{q}\mid
X^{\sigma 0}_{q}\rangle\!\rangle_{\omega},
\label{30}
\end{equation}
is a single-particle spectral function
which does not depend on spin $\sigma$ in the paramagnetic state.
 We have also introduced the unified spin-charge fluctuation spectrum
\begin{equation}
  \chi_{cs}^{\prime\prime}({\bf q},\omega)
=
\frac{1}{4\pi}N_{\bf q}^{\prime\prime}(\omega)
+\chi_{\bf q}^{\prime\prime}(\omega)
,
\label{31}
\end{equation}
with 
\begin{equation} 
\chi_{\bf q}(\omega)=-\frac{1}{\pi}\langle\langle 
{\bf S}_{\bf q}|{\bf S}_{-{\bf q}}\rangle\rangle_{\omega}
\label{32}
\end{equation}
being the dynamic spin susceptibility. 
In obtaining Eq.(\ref{29})
we have also used the identity
$\langle\langle  S_{\bf q}^{\sigma}| 
S_{-{\bf q}}^{\bar{\sigma}}\rangle\rangle_{\omega}=2
\langle\langle  S_{\bf q}^{z}| 
S_{-{\bf q}}^{z}\rangle\rangle_{\omega}$
which holds in the paramagnetic state.

As it is clear from Eq.(\ref{27}) the memory function involves two
contributions; the first one (\ref{28}) stems from the $H_{t}$ term (\ref{1})
and describes the p-h contribution, while the second one comes
both from $H_{t}$ and $H_{J}$ parts and involves electron scattering 
on spin and charge fluctuations. Further  we assume that the charge carriers 
are mainly relaxed by the scattering on spin fluctuations  
 and retain only the second term  $\chi_{\bf q}^{\prime\prime}(\omega)$
in Eq.(\ref{31}).

For further discussion it is more convenient to integrate out
the fermionic degrees of freedom $({\bf k},\omega_{2})$ in 
$\Pi_{2}^{\prime\prime}({\bf q},\omega)$ (\ref{29}), which
 results
\begin{eqnarray}
\Pi_{2}^{\prime\prime}({\bf q},\omega)&=&\frac{1}{N}
\sum_{\bf p}\int\limits_{-\infty }^{\infty }
d\omega_{1}\left[1+N_{\omega_{1}}+N_{\omega-\omega_{1}}\right]\nonumber\\
&\times& \tilde{\Pi}^{\prime\prime}_{{\bf q}-{\bf p}}(\omega-\omega_{1})
\chi^{\prime\prime}
_{ \bf p}(\omega_{1}),
\label{33}
\end{eqnarray}
where we have introduced an effective spectral function
\begin{eqnarray}
\tilde{\Pi}^{\prime\prime}_{{\bf q}-{\bf p}}(\omega)&=&
\frac{-2\pi}{N}
\sum_{\bf k}M^{2}_{{\bf k},{\bf q},{\bf p}}
\int\limits_{-\infty }^{\infty }
d\omega_{1}n_{\omega_{1},\omega}\nonumber\\
&\times& 
A_{{\bf k}+{\bf q}-{\bf p}}(\omega+\omega_{1})
A_{{\bf k}}(\omega_{1})
\label{34},
\end{eqnarray}
for particle-hole excitations coupled to a particular $({\bf p},\omega)$
state of spin fluctuations.

To conclude the section, we calculate the first moment of DCS (\ref{15}).
By performing the commutation between the density and ``current'' (\ref{11})
operators we readily get
\begin{equation}
m_{\bf q}=4ztN_{1}(1-\gamma_{\bf q}),
\label{35}
\end{equation}
where 
\begin{equation}
N_{m}=\frac{1}{N}\sum_{\bf q}\gamma_{\bf q}^{m}
\langle X_{\bf q}^{\sigma 0}X_{\bf q}^{0\sigma}\rangle 
\label{36}
\end{equation}
is the p-h corelation function.
\section{Dynamic charge susceptibility}
Equations (\ref{8}),(\ref{9}), and (\ref{12}) result in the following
form of DCS
\begin{equation}
N_{\bf q}(\omega)=-\frac{m_{\bf q}}
{\omega^{2}- \left[\Pi({\bf q},\omega)-\Pi({\bf q},0)\right]/m_{\bf q}
-\Omega_{\bf q}^{2}},
\label{37}
\end{equation}
where $m_{\bf q}$ is the first moment of DCS given by Eq.(\ref{35}),
and $\Omega_{\bf q}^{2}=m_{\bf q}/N_{\bf q}$  is a  mean field (MF) 
spectrum for the density fluctuations. The memory function formalism does not
provide itself the static susceptibility $N_{\bf q}$. The latter
is calculated within the same approximation scheme as for 
the static spin susceptibility, \cite{jackeli}
which results in the following form of the MF spectrum
\begin{equation}
\Omega_{\bf q}^{2}=2z^{2}t^{2}C_{\bf q}(1-\gamma_{\bf q})
\label{38}
\end{equation}
where
\begin{equation}
C_{\bf q}=\frac{1}{z}(1-\frac{n}{2})+N_{2}-\frac{J}{2zt}N_{1}(1+z\gamma_{\bf q})
\label{39}
\end{equation}
and the parameters $N_{m}$ are defined by Eq.(\ref{36}).
Here we note that the MF spectrum $\Omega_{\bf q}$
resembles the dispersion of an undamped collective mode in the charge channel
found in the leading order of $1/N$ expansion. \cite{wang}

In a proper analysis, Eqs. (\ref{27})-(\ref{29}) and (\ref{37})
should be treated self-consistently with the equations for the single--particle
spectral function $A_{\bf k}(\omega)$ \cite{plakida} and the 
spin susceptibility
$\chi_{\bf q}(\omega)$. \cite{jackeli} This problem, to our knowledge,
can be solved only numerically. Here we  show, that  the main
features of charge fluctuation spectrum observed 
in the exact diagonalization studies \cite{tohoyama}
can be, at least qualitatively, reproduced in an analytic way based on the
physically justified  anzats for the single--particle spectral function and 
the dynamic spin susceptibility. First, we discuss the one-particle
spectral function $A_{\bf k}(\omega)$.

Actually,  the spectral characteristics of the $t-J$ model
 have been investigated  
by  various analytic and numerical approaches.
\cite{daggoto} Those results led to the consensus that the single--particle
 spectrum involves a narrow quasiparticle (QP) band of coherent states 
and a broad continuum of incoherent states. The corresponding spectral function can be represented as
\begin{equation}
A_{{\bf k}}( \omega)=A^{\text{coh}}_{{\bf k}}(\omega)+
A^{\text{inc}}_{{\bf k}}(\omega), 
\label{40}
\end{equation}
where the QP part is given by
\begin{equation}
A^{\text{coh}}_{{\bf k}}(\omega)=Z_{{\bf k}}\delta
(\omega-\epsilon_{{\bf k}}),
\label{41}
\end{equation}
with $Z_{{\bf k}}$ and $\epsilon_{{\bf k}}=\varepsilon_{{\bf k}}-\mu$ being the
QP weight  and dispersion referred to the chemical potential $\mu$, 
respectively. While the incoherent part
$A^{\text{inc}}_{{\bf k}}(\omega)$ is little affected by the doping,
the coherent band structure strongly depends on the magnetic background.
Namely, in the low doping regime (ordered phase) the QP dispersion
is determined by the hopping within a given AFM sublattice \cite{daggoto}
and the Fermi  surface (FS) consists of  small hole--pockets 
centered around $(\pm \pi/2,\pm \pi/2)$.
While for a moderate doping (paramagnetic state) the dispersion reflects the
dominance of n.n. hopping, there exists large electronic FS which
encloses a fraction of the Brillouin zone equal to the electron
concentration $n$. \cite{stephan} 
For the latter doping regime, the exact diagonalization results are well
fitted by a simple tight-binding dispersion with some
effective hopping amplitude $t_{\text{eff}}$ 
which scales with $J$ ( for $\delta=0.1$ and $J=0.4t$
 $t_{\text{eff}}=0.24t$ \cite{stephan}).
 Hence,
in the paramagnetic phase we can put
$\varepsilon_{{\bf k}}=-zt_{\text{eff}}\gamma_{{\bf k}}$.

A nearly structureless incoherent part is predominantly distributed below 
the
QP band (in the electronic picture) and we approximate 
$A^{\text{inc}}_{{\bf k}}(\omega)$ as follows
\begin{equation}
A^{\text{inc}}_{{\bf k}}(\omega^{\prime})=\frac{1}{\Gamma}
\theta(-\omega^{\prime})\theta (W_{\text{inc}}+\omega^{\prime})
,
\label{42}
\end{equation}
where $W_{{\text{inc}}}\simeq 5t$ 
is the incoherent bandwidth and $\omega^{\prime}$
is measured from the bottom of the QP band. The spectral weights of the
incoherent  band $\Gamma$ and QP $Z_{\bf k}$ are provided by  the
sum rules
\begin{eqnarray} 
&&\frac{1}{N}\sum_{\bf k}\int\limits_{-\infty }^{\infty }d\omega
 A_{{\bf k}}(\omega)=1-\frac{n}{2},\nonumber\\
&&\frac{1}{N}\sum_{{\bf k},\sigma}\int\limits_{-\infty }^{0 }d\omega
 A_{{\bf k}}(\omega)=n,
\label{43}
\end{eqnarray}
and are given by
\begin{equation}
Z=\frac{2\delta}{1+\delta},\;\;\;\;
\Gamma=\frac{2(1+\delta)}{(1-\delta)^{2}}W_{{\text{inc}}},
\label{44}
\end{equation}
where $Z=\langle Z_{\bf k}\rangle$ is the averaged QP weight and it
coincides with that obtained 
within the Gutzwiller approximation. \cite{vollhardt}

With  the above form of the spectral function (\ref{40})-(\ref{42}),
 the equal-time corelation functions $N_{1,2}$ (\ref{39}) 
are estimated to be
\begin{equation}
N_{1}\simeq\frac{2}{\pi^{2}}nZ,\;\;\;N_{2}\simeq\frac{n}{2z}
\label{45}
\end{equation}

Finally, we assume that the spin susceptibility $\chi_{\bf q}(\omega)$ (\ref{32})
is peaked at the AFM wave vector ${\bf Q}=(\pi,\pi)$. \cite{incom}
Then, at $T=0$, we approximate $\Pi_{2}^{\prime\prime}({\bf q},\omega)$ (\ref{33}) as follows
\begin{eqnarray}
\Pi_{2}^{\prime\prime}({\bf q},\omega)&\simeq&\int\limits_{0 }^{\omega}
d\omega_{1}
 \tilde{\Pi}^{\prime\prime}_{{\bf q}-{\bf Q}}(\omega-\omega_{1})
\chi^{\prime\prime}(\omega_{1}),
\label{46}
\end{eqnarray}
where 
\begin{equation}
\chi(\omega)=\frac{1}{N}\sum_{\bf p}\chi_{\bf p}^{\prime\prime}(\omega)
\label{47}
\end{equation}
 is the local spin susceptibility. 
Since the detailed form of  $\chi(\omega)$
is not essential for our study,  we use the following relaxation-type 
susceptibility
\begin{equation}
\chi(\omega)=\frac{\chi}{1-i\omega/\omega_{\text{0}}}
\label{48}
\end{equation}
which seems to be in agreement with the exact diagonalization data.
\cite{tohoyama,prelov} In Eq.(\ref{48}) $\omega_{\text{0}}\propto J$  
 is the energy scale of spin fluctuations 
and $\chi$ is the static susceptibility provided by the
sum rule
\begin{equation} 
\int\limits_{0}^{\omega_{\text{c}}}d\omega
 \chi(\omega)=\frac{3}{4}n,
\label{49}
\end{equation}
with the high-energy cutoff $\omega_{\text{c}}=2J$.
Equations (\ref{48}) and (\ref{49}) result in
\begin{equation} 
\chi=\frac{3n}{2\omega_{\text{0}}\ln[1+
(\omega_{\text{c}}/\omega_{\text{0}})^2]}.
\label{50}
\end{equation}

\subsection{Long--wavelength limit}

First, we discuss a small ${\bf q},\omega$ limit of DCS.
In the long wave-length limit, $q\rightarrow 0$, the MF spectrum (\ref{38})
reduces  to the sound mode $\Omega_{\bf q}=v_{\text{s}}q$ with 
the sound velocity  $v_{\text{s}}=zt\sqrt{C_{0}/2}$ larger than the
 Fermi velocity $v_{\text{F}}(\mu)=zt_{\text{eff}}\sqrt{(1-\tilde{\mu}^2)/2}$
( $\tilde{\mu}=|\mu|/zt_{\text{eff}}\simeq\pi\delta/4$
for small $\delta$).

At small  $q$ the vertex functions (\ref{21})-(\ref{24}) 
are given by 
\begin{eqnarray}
G_{{\bf k},{\bf q}}&=&2t^{2}[2z(\hat{\bf q}\nabla_{\bf k}\gamma_{\bf k})^{2}
-1]q^{2},
\label{51}\\
M_{{\bf k},{\bf Q},{\bf p}}&=&z(z-1)tJ
(\hat{\bf q}\nabla_{\bf k}\gamma_{\bf k})q.
\label{52}
\end{eqnarray}
with $\hat{\bf q}={\bf q}/q$.

First we consider the real part of ``self--energy''
$\Pi({\bf q},\omega)$ (\ref{27})--(\ref{29}).
Since for small $q$, the vertex functions
$G_{{\bf k},{\bf q}}\sim q^{2}$
and $M_{{\bf k},{\bf Q},{\bf p}}\sim q$,
we approximate $\Pi^{\prime}({\bf q},\omega)\simeq\Pi_{2}^{\prime}({\bf q},\omega)$
to keep the contributions  leading in small $q$.
Moreover, for small $\omega$ we expand $\Pi_{2}^{\prime}({\bf q},\omega)$
as  
\begin{equation}
\Pi_{2}^{\prime}({\bf q},\omega)\simeq\Pi_{2}^{\prime}({\bf q},0)
-\alpha_{\bf q}\omega^{2},
\label{53} 
\end{equation}
where $\alpha_{\bf q}>0$ and is given by
\begin{equation}
\alpha_{\bf q}=-\frac{1}{2}\frac{d^{2}\Pi_{2}^{\prime}({\bf q},\omega)}
{d\omega^{2}}\biggl|_{\omega=0}=-\frac{1}{\pi}\int\limits_{-\infty }^{\infty }
\frac{\Pi_{2}^{\prime\prime}({\bf q},\omega)}{\omega^{3}}d\omega,
\label{54} 
\end{equation}
Since $\Pi_{2}^{\prime\prime}({\bf q},\omega)$ (\ref{46}) is an odd function
of $\omega$, there is no term linear in $\omega$  in the expansion
(\ref{53}).
Equations (\ref{37}) and (\ref{53}) result in the following form of DCS
for small $q,\omega$
\begin{equation}
N_{\bf q}(\omega) \simeq\frac{ - m_{\bf q}/(1+\lambda)}
{\omega^{2}-\tilde{v}_{s}^{2}q^{2}+2i\omega\Gamma_{\bf q}}, 
\label{55}
\end{equation}
where 
\begin{equation}
\lambda=\lim\limits_{{\bf q}\to 0}\frac{\alpha_{\bf q}}{m_{\bf q}},\;
\tilde{v}_{s}=\frac{v_{s}}{\sqrt{1+\lambda}},\;
\Gamma_{\bf q}=\frac{-\Pi^{\prime\prime}({\bf q},\omega )}{2m_{\bf
q}(1+\lambda)\omega},
\label{56}
\end{equation}
$\tilde{v}_{s}$ and $\Gamma_{\bf q}$
are the renormalized sound velocity and sound damping, respectively.

First, we consider  
renormalization factor $\lambda$.
The estimation of $\lambda$  is interesting by itself, 
since it represents the  electron mass enhancement factor [see Sec. V]
  and can be related to experiment.
\cite{puchkov} 
Approximation (\ref{40}) for the one-particle spectral
function $A_{\bf k}(\omega)$ leads to  two different contributions to
the effective spectral function for p-h excitations
$\tilde{\Pi}^{\prime\prime}_{{\bf q}-{\bf Q}}(\omega)$ (\ref{34}).
The first one 
$\tilde{\Pi}^{\prime\prime}_{{\bf q}-{\bf Q}}(\omega)^{\text{c-c}}$
 is due to the transitions within the QP band and
the remaining part
$\tilde{\Pi}^{\prime\prime}_{{\bf q}-{\bf Q}}(\omega)^{\text{i-c}}$
is provided by the incoherent-coherent transitions.

First, considering 
$\tilde{\Pi}^{\prime\prime}_{{\bf q}-{\bf p}}(\omega)^{\text{c-c}}$
(\ref{34}), (\ref{51})
we come to the following expression for small $q$
\begin{eqnarray}
\tilde{\Pi}^{\prime\prime}_{{\bf q}-{\bf Q}}(\omega)^{\text{c-c}}&=&
\frac{-2\pi\Lambda Z^{2}}{N}q^{2}
\sum_{\bf k}(\hat{\bf q}\nabla_{\bf k}\gamma_{\bf k})^{2}\nonumber\\
\times\!\!\!&[&\!\!\!n(\epsilon_{\bf k})-n(\epsilon_{\bf k}+\omega)]
\delta(\omega-\epsilon_{{\bf k}+{\bf q}-{\bf Q}}+\epsilon_{\bf k})
\label{57},
\end{eqnarray}
with $\Lambda=[z(z-1)tJ]^{2}$. As it has been previously discussed by several authors. \cite{benard}
the p-h spectral function for tight-binding electrons
exhibits a crossover at frequency $\omega=2|\mu|$.
Namely, for $\omega<2|\mu|$ it is peaked
 at the incommensurate wave vectors ${\bf Q}_{*}=(\pi\pm \delta^{*},\pi),
(\pi,\pi\pm \delta^{*})$ where the displacement $\delta^{*}$ 
for small $\mu$ is  
given by $\delta^{*}\simeq\mu/t_{\text{eff}}$, whereas for
$\omega >2|\mu|$  the p-h spectral function gets its maximum value
at AFM wave vector 
${\bf Q}$ and follows the nested Fermi liquid scaling. \cite{virosztek}
Hence we consider the cases $\omega <2|\mu|$ and $\omega >2|\mu|$
separately. 

In the case $\omega < 2|\mu|$  we  put 
$\epsilon_{{\bf k}+{\bf q}-{\bf Q}}=\epsilon_{{\bf k}-{\bf Q}_{*}}$ in
Eq.(\ref{57}) and 
for $\delta^{*}\ll 1$ expand 
$\epsilon_{{\bf k}+{\bf Q}_{*}}\simeq-
\epsilon_{\bf k}-\delta^{*}\partial\epsilon_{\bf k}/\partial k_{x}$.
Moreover, since the dominant contribution to the integral in Eq.(\ref{54})
comes from $\omega \sim 0$,  at $T=0$ we approximate
$n(\epsilon_{\bf k})-n(\epsilon_{{\bf k}+\omega})
\simeq\omega\delta(\epsilon_{\bf k})$
and, as a result, we get
\begin{eqnarray}
\tilde{\Pi}^{\prime\prime}_{{\bf q}-{\bf Q}}(\omega)^{\text{c-c}}&=&
\frac{-2\pi\Lambda tZ^{2}}{W_{\text{coh}}^{2}|\mu|}q^{2}\omega I_{1}(\omega)
\theta(2|\mu|-\omega)
\label{57a},
\end{eqnarray}
where  $W_{\text{coh}}=2zt_{\text{eff}}$ is the coherent bandwidth, and
$$
I_{1}(\omega)=\frac{2\sin^{2}k^{*}_{x}}{\pi^{2}|\cos k^{*}_{x}\sin k^{*}_{y}|}
$$
with $\sin k^{*}_{x}=1-\omega/(2|\mu|)$ 
and $\cos k^{*}_{y}=2\tilde{\mu}-\cos k^{*}_{x}$. For 
$\omega\ll 2|\mu|$ we have $I_{1}(\omega)\simeq2\sqrt{\omega/|\mu|}/\pi^{2}$
and thus
\begin{eqnarray}
\tilde{\Pi}^{\prime\prime}_{{\bf q}-{\bf Q}}(\omega)^{\text{c-c}}\simeq
\frac{-4\Lambda tZ^{2}}{\pi W_{\text{coh}}^{2}}\sqrt{\frac{\omega}{|\mu|}}
q^{2}
\label{57b}.
\end{eqnarray} 
Then from Eqs. (\ref{46}) and (\ref{57b}) for the velocity renormalization factor
$\lambda$ (\ref{56}) we get
\begin{eqnarray}
\lambda^{c-c}_{1}\simeq
\frac{\sqrt{2}\Lambda Z^{2}\chi}{15\pi^{2} zt_{\text{eff}}^{2}N_{1}\omega_{\text{0}}}
\label{58}
\end{eqnarray} 
where the QP weight $Z$ and n.n p-h correlator $N_{1}$ 
are given by Eqs.(\ref{44}) and (\ref{45}).
For the actual values of the parameters $J=0.4t$, $t_{\text{eff}}=0.24t$, and $\delta=0.1$ we obtain $\lambda^{c-c}_{1}\simeq 4$.

Next we consider the case $\omega>2|\mu|$. 
In this case the  p-h spectral
 function is peaked at the  AFM 
wave vector $Q$ and from Eq.(\ref{57}) we come to
\begin{eqnarray}
\tilde{\Pi}^{\prime\prime}_{{\bf q}-{\bf Q}}(\omega)^{\text{c-c}}=
\frac{-\pi\Lambda Z^{2}}{W_{\text{coh}}}q^{2}I(\tilde{\omega})
\theta(\tilde{\omega}
-2\tilde{\mu})\theta(2-\tilde{\omega})
\label{59},
\end{eqnarray}
where $\tilde{\omega}=2\omega/W_{\text{coh}}$ and 
\begin{eqnarray}
I(\tilde{\omega})&=&\frac{1}{\pi^{2}}\left[
\tilde{\omega}_{+}E(\tilde{\omega}_{-}/\tilde{\omega}_{+})-
2\tilde{\omega}K(\tilde{\omega}_{-}/\tilde{\omega}_{+})\right]
\label{60}
\end{eqnarray}
with $\tilde{\omega}_{\pm}=2\pm\tilde{\omega}$, $K(x)$ and $E(x)$
are the complete elliptic integrals of the first and second kind,
respectively.
The function $I(x)$ (\ref{60}) is normalized to $1/4$ 
in the interval $0<x<2$ and well approximated by the following
linear dependence $I(x)=(2-x)/8$ and we get
\begin{eqnarray}
&\mbox{}&\tilde{\Pi}^{\prime\prime}_{{\bf q}-{\bf Q}}(\omega)^{\text{c-c}}\simeq
\frac{-\pi\Lambda Z^{2}}{4W^{2}_{\text{coh}}}f(\omega)q^{2}\nonumber\\
&\mbox{}&f(\omega)=(W_{\text{coh}}-\omega)
\theta(\omega
-2|\mu|)\theta(W_{\text{coh}}-\omega).
\label{61}
\end{eqnarray}
That results in the following  contribution 
 to $\lambda$ 
\begin{equation}
\lambda^{\text{c-c}}_{2}\simeq\frac{2(z-1)^{2}tJZ^{2}}{N_{1}W_{\text{coh}}^{2}}I,
\label{62}
\end{equation}
with
\begin{equation}
I=\int\limits_{0 }^{\infty }\int\limits_{0 }^{x}dxdy
\frac{\tilde{f}(x-y)}{x^{3}}
\tilde{\chi}(y)\theta(2-y)
\label{63}
\end{equation}
where the dimensionless functions
$\tilde{f}(x)$ and $\tilde{\chi}(y)$ stand for
$f(\omega)$ (\ref{61}) and spin susceptibility $\chi(\omega)$ (\ref{48})
measured in units of $J$. 
For 
$J=0.4t$ and $\delta=0.1$ we  calculated integral (\ref{63})
numerically
and found $I=0.5$. Then from Eq.(\ref{62}) we estimate
$\lambda^{\text{c-c}}_{2}\simeq 0.8$.

As for  the remaining contribution
$\tilde{\Pi}^{\prime\prime}_{{\bf q}-{\bf Q}}(\omega)^{\text{i-c}}$,
with the help of Eqs.(\ref{34}),(\ref{42}), and (\ref{52}) 
at $T=0$ we obtain
\begin{eqnarray}
\tilde{\Pi}^{\prime\prime}_{{\bf q}-{\bf Q}}(\omega)^{\text{i-c}}&\simeq&
\frac{-\pi\Lambda Z}{4z\Gamma}q^{2}\nonumber\\
&\times&\theta(2\omega-W_{\text{coh}}+2|\mu|)
\theta(W_{\text{tot}}-\omega),
\label{64}
\end{eqnarray}
where $W_{\text{tot}}=W_{\text{coh}}+W_{\text{inc}}$ is the total bandwidth.
From Eq.(\ref{56}) and by using the sum rule (\ref{49}),
the upper value of the incoherent-coherent contributions
to  $\Pi_{2}^{\prime\prime}({\bf q},\omega)^{\text{i-c}}$
is estimated as
\begin{equation}
\Pi_{2}^{\prime\prime}({\bf q},\omega)^{\text{i-c}}\simeq
\frac{3}{4}n\tilde{\Pi}^{\prime\prime}_{{\bf q}-{\bf Q}}(\omega)^{\text{i-c}}.
\label{65}
\end{equation}
That results in
\begin{equation}
\lambda^{i-c}\simeq
\frac{3n(z-1)^{2}ZtJ^{2}}{16\Gamma N_{1}W_{\text{coh}}^{2}}
\left[4-\frac{W_{\text{coh}}^{2}}{W_{\text{tot}}^{2}}\right]\simeq 0.1.
\label{66}
\end{equation}

A small value of $\lambda^{i-c}\simeq 0.1$ in comparison with
 $\lambda^{i-c}\simeq4$
is due to a large threshold energy
($W_{\text{coh}}/2-|\mu|\simeq t)$ (\ref{64}) for  incoherent-coherent
transitions which are important only
in describing  high-energy density fluctuations.  

Finally, by summing all three contributions (\ref{60}), (\ref{64}), and (\ref{68})
for the velocity renormalization factor we get $\lambda\simeq 5$.
Here we notice, that in the  Fermi liquid theory, for
$v_{\text{s}}>v_{\text{F}}$  ``self-energy'' corrections  stiffen
the sound. \cite{pines} Contrary to this,
, in the present case  the sound softens.
That is due to
the scattering on spin fluctuations given by the $\Pi_{2}({\bf q},\omega)$ 
term (\ref{46}).
One can easily verify that the renormalized sound velocity (\ref{56})  gets   smaller
than the Fermi one $\tilde{v_{\text{s}}}=v_{\text{s}}/\sqrt{1+\lambda}
<v_{\text{F}}$.
It falls down into a particle-hole continuum getting
 finite damping due to the decay  into particle-hole pairs. \cite{remark}
This process is described by $\Pi_{1}^{\prime\prime}({\bf q},\omega)$
(\ref{28}) .
 The latter  in the small $q,\omega$ limit
reads as [see Eqs.(\ref{28}) and (\ref{51})] 
\begin{eqnarray}
\Pi_{1}^{\prime\prime}({\bf q},\omega)&=&
\frac{-8\pi t^{4}Z^{2}}{N}\omega q^{3}
\sum_{\bf k}[2z(\hat{\bf q}\nabla_{\bf k}\gamma_{\bf k})^{2}-1]^{2}\nonumber\\
&\times&
\delta(\epsilon_{\bf k})\delta(\omega/q-\hat{\bf q}{\bf v}_{\bf k}),
\label{67}
\end{eqnarray} 
where ${\bf v}_{\bf k}=\nabla_{\bf k}\epsilon_{\bf k}$ is the QP velocity.
The integration over ${\bf k}$ in Eq.(\ref{67}) results in
\begin{eqnarray}
\Pi_{1}^{\prime\prime}({\bf q},\omega)=
\frac{-\sqrt{2}t^{4}Z^{2}}{\pi t_{\text{eff}}^{2}|\sin\theta|}
\left[1-\left(\frac{2\omega}{qv_{\text{F}}}\right)^{2}\right]
\omega q^{3},
\label{68}
\end{eqnarray} 
where $\cos\theta=[\mu^{2}+2\omega^{2}/q^{2}]/v_{\text{F}}^{2}-1$.
This gives the following form of the sound damping 
\begin{eqnarray}
\Gamma_{q}\simeq\beta \tilde{v}_{\text{s}}q\;\;\;\;
\beta=\frac{Z^{2}t^{3}}{\pi zt_{\text{eff}} N_{1}v_{\text{s}}^{2}},
\label{69}
\end{eqnarray} 
for actual values of the parameters  $\beta <1$, 
and thus, in accordance with Ref.~\onlinecite{horsch},
one obtains that the sound damping is only numerically smaller
than its energy.

\subsection{Short--wavelength limit}
At large momenta the main spectral weight of density fluctuations
is located at high energies, $(\sim t)$, near the MF spectrum   
$\Omega_{\bf q}$ (\ref{38}).
For instance, at ${\bf q}={\bf Q}$ we have $\Omega_{\bf Q}\simeq zt$
while in the exact diagonalization studies \cite{tohoyama}
the peak is observed at $\omega\simeq 6t$.
However, considering  `` self-energy'' corrections and noting
that $\Pi^{\prime}_{\bf q}(\omega)$ falls off as $1/\omega^{2}$
at large frequencies, from Eq.(\ref{37})
we obtain the renormalized spectrum as
$\tilde{\Omega}_{\bf q}\simeq \sqrt{\Omega_{\bf q}^{2}-\Pi^{\prime}_{\bf q}
(0)/m_{\bf q}}$. 
One can easily show that $\Pi^{\prime}_{\bf q}(0)<0$ and hence
the  spectrum is shifted to higher energies.
At large momenta the peak is dispersed out 
of the coherent p-h continuum and 
its broadening is only due to high-energy transitions involving the
incoherent band. Since the latter  has been neglected 
in Ref.~\onlinecite{zeyher}, the authors observed an infinitely sharp peak.
However, as it follows, the damping of the high energy mode is
comparable to its energy. 
Near the pole $\tilde{\Omega}_{\bf q}$  we estimate the damping as
\begin{equation}
\tilde{\Gamma}_{\bf q}
=\frac{-\Pi^{\prime\prime}({\bf q},\tilde{\Omega}_{\bf q})}{2m_{\bf
q}\tilde{\Omega}_{\bf q}}
\label{70}
\end{equation}
where $\Pi^{\prime\prime}({\bf q},\omega)=
\Pi^{\prime\prime}_{1}({\bf q},\omega)^{\text{i-c}}+
\Pi^{\prime\prime}_{2}({\bf q},\omega)^{\text{i-c}}$ 
describes incoherent-coherent transitions. From Eqs.(\ref{28}) and
(\ref{42}) for 
${\bf q}={\bf Q}$  we obtain
\begin{equation}
\Pi^{\prime\prime}_{1}({\bf Q},\omega)^{\text{i-c}}\simeq
\frac{-5\pi(zt)^{4} Z}{4\Gamma}.
\label{71}
\end{equation}
The second contribution 
$\Pi^{\prime\prime}_{2}({\bf q},\omega)^{\text{i-c}}$ from Eqs.(\ref{34}),
(\ref{42}), and (\ref{65}) is estimated as
\begin{equation}
\Pi^{\prime\prime}_{2}({\bf Q},\omega)^{\text{i-c}}\simeq
\frac{-6n\pi(zt)^{4} Z}{5\Gamma}.
\label{72}
\end{equation} 
Equations (\ref{70}),(\ref{71}),  and (\ref{72}) result in
\begin{equation}
\tilde{\Gamma}_{\bf q}\simeq\frac{2(25+24n)\pi t^{3} Z}{5N_{1}
\Gamma\tilde{\Omega}
_{\bf q}}\sim 3t.
\label{73}
\end{equation} 
Thus, the  peak gets rather broad in accordance
with the exact diagonalization results. \cite{tohoyama}

For large momenta but low energies, charge excitation spectrum should show
some low energy structure related to the contribution from
the  p-h continuum
to $N_{\bf q}^{\prime\prime}(\omega)$. 
Since  $\Omega_{\bf q}$
is larger in $(\xi,\xi)$ direction than in $(\xi,0)$, 
the low energy structure should be less pronounced in the latter case.
The same anisotropy
 has been observed in the exact diagonalization studies.
\cite{tohoyama}

\section{Optical conductivity}
In this section, we discuss  the optical conductivity $\sigma(\omega)$.
In the linear response theory of Kubo \cite{kubo}, the frequency-dependent
conductivity is given by the relaxation function for
currents
\begin{equation}
\sigma_{xx}(\omega)=\frac{ie^{2}}{V}((J_{x}|J_{x}))_{\omega}.
\label{74}
\end{equation}
By using the continuity equation and equation of motion for the
GF's, one can easily relate the longitudinal conductivity to the dynamic
charge  susceptibility (\ref{5})
\begin{equation}
\sigma_{xx}(\omega)=-\frac{ie^{2}}{V}\lim\limits_{ q \to 0}
\frac{\omega N_{\bf q}(\omega)}{q^{2}},
\label{75}
\end{equation}
where $q=q_{x}$. From Eqs. (\ref{37}) and (\ref{75}) we express conductivity in terms of the
memory function
\begin{equation}
\sigma_{xx}(\omega)=\frac{ie^{2}}{V}
\frac{D}{\omega-M(\omega)}
\label{76}
\end{equation}
where $D=\lim\limits_{ q \to 0}m_{\bf q}/q^{2}=ztN_{1}$ is the Drude weight
which is given by one-half  the averaged kinetic energy
 $D=-\langle H_{t}\rangle$/2, $m_{\bf q}$ and $N_{1}$ are defined by Eqs. 
(\ref{35}) and (\ref{36}), respectively. 
The memory function $M(\omega)$
reads as
\begin{equation}
M(\omega)=
\frac{\Pi ( \omega)-
\Pi ( 0)}{\omega },
\label{77}
\end{equation}
where 
\begin{equation}
\Pi(\omega)=\lim\limits_{ q \to 0}\frac{\Pi ({\bf q}, \omega)}{D q^{2}},
\label{78}
\end{equation}
with $\Pi ({\bf q}, \omega)$ defined by Eq. (\ref{27}). Since for
small $q$, $\Pi_{1}({\bf q}, \omega)\sim q ^{4}$ and
$\Pi_{2}({\bf q}, \omega)\sim q^{2}$, 
 only the second  one 
contributes to $\Pi(\omega)$. The  latter  is  given by 
Eqs.(\ref{33})and (\ref{34})  at $q=0$ with 
$M_{{\bf k},{\bf q}, {\bf p}}$ replaced by the transport vertex given by
\begin{eqnarray}
M_{{\bf k},{\bf p}}=\lim\limits_{ q \to 0}\frac{M_{{\bf k},{\bf q}, {\bf p}}}
{q^{2}}&=&[t_{{\bf k}-{\bf p}}v_{\bf k}-t_{\bf k}v_{{\bf k}-{\bf p}}
-2tv_{\bf p}]\nonumber\\
&+&\frac{1}{2}[J+J_{\bf p}][v_{{\bf k}-{\bf p}}-v_{\bf k}],
\label{79}
\end{eqnarray}
where $t_{\bf k}=zt\gamma_{\bf k}$, $J_{\bf k}=zJ\gamma_{\bf k}$, and
$v_{\bf k}=\partial t_{\bf k}/\partial k_{x}$. 

We rewrite  conductivity (\ref{76}) in the form of the generalized Drude law
as follows
\begin{equation}
\sigma_{xx}(\omega)=\frac{e^{2}}{V}
\frac{\tilde{D}(\omega)}{1/\tilde{\tau}(\omega)-i\omega},
\label{80}
\end{equation}
where an effective Drude weight and the relaxation time are given by
\begin{equation}
\tilde{D}(\omega)=\frac{D}{1+\lambda(\omega)}\;\;\;
\frac{1}{\tilde{\tau}(\omega)}=\frac{1}{\tau(\omega)(1+\lambda(\omega))},
\label{81}
\end{equation}
 with
\begin{equation}
\lambda(\omega)=-\frac{M^{\prime}(\omega)}{\omega}\;\;\;
\frac{1}{\tau(\omega)}=-M^{\prime\prime}(\omega),
\label{82}
\end{equation}
and $1+\lambda(\omega)$ is the interaction-induced optical 
mass enhancement factor.
The latter  in the static limit is calculated in the preceding section
 and is estimated to be of order 6.
 That is in a
good  agreement with the optical measurement data.
\cite{puchkov} Optical conductivity of the $t-J$ model within the present formalism was studied by one of us (N.P.) in Ref.\onlinecite{plakida1},
where temperature and frequency dependence of 
$\sigma(\omega)$ where discussed.
Here we  mainly focus on the analysis of the 
low-frequency behavior of the relaxation rate 
\begin{equation}
\Gamma(\omega)=\frac{1}{\tau(\omega)}=-
\frac{\Pi^{\prime\prime}(\omega)}{\omega}.
\label{83}
\end{equation}

Following  Sec. IV  we approximate 
$\Pi^{\prime\prime}(\omega)$ as
\begin{eqnarray}
\Pi^{\prime\prime}(\omega)\simeq\int\limits_{0 }^{\omega}\!\!
d\omega_{1}
\chi^{\prime\prime}(\omega_{1})\times\left\{
\begin{array}{ll}
\tilde{\Pi}^{\prime\prime}_{\bf Q_{*}}(\omega-\omega_{1})& \omega<2|\mu|\\
\tilde{\Pi}^{\prime\prime}_{\bf Q}(\omega-\omega_{1})&  \omega>2|\mu|\\
\end{array}\right.
\label{im}
\end{eqnarray}
 In the case $\omega<2|\mu|$, the effective spectral function of p-h excitations $\tilde{\Pi}^{\prime\prime}_{\bf Q_{*}}(\omega)$
(\ref{57b})
 is given by 
\begin{equation}
\tilde{\Pi}^{\prime\prime}_{{\bf Q}_{*}}(\omega)\simeq
\frac{-4\Lambda tZ^{2}}
{\pi W_{\text{coh}}^{2}D}\sqrt{\frac{\omega}{|\mu|}}.
\label{88}
\end{equation}
We remark the square-root behavior of the p-h spectral function 
$\Pi^{\prime\prime}_{{\bf Q}_{*}}(\omega) \sim \sqrt{\omega}$ instead of the 
conventional linear $\omega$-dependence. \cite{pines}
This behavior results  in  the square-root singularity of the
structure factor that is known as $2k_{\text{F}}$ anomaly
familiar for the electron system in low dimensions. \cite{benard}
It also leads to the deviation from the conventional square law resulting
  in the following form of the relaxation rate
\begin{equation} 
\Gamma(\omega)\simeq
\frac{16\Lambda Z^{2}t\chi}{15\pi D W_{\text{coh}}^{2}\sqrt{|\mu|}
\omega_{\text{0}}}
\omega^{3/2}.
\label{89}
\end{equation}
Here we note that the $\omega^{3/2}$-law
of inverse life time for electron states near the saddle points  
has been obtained in Refs.~\onlinecite{gopalan,rice}. In the present case,
the Van Hove singularity plays no role. 
The obtained $\omega^{3/2}$--dependence of the relaxation 
time is rather due to the  coexistence  of the
 peak in the spin fluctuation spectrum and the
$2k_{\text{F}}$ ``anomaly'' in the p-h spectral function
at $q\sim Q$. We notice
that the former one, $2k_{\text{F}}$ ``anomaly'', is not related to the
FS topology and is inherent in a low-dimensional electron system.

Now we consider the region  $\omega > 2|\mu|$. In this case the p-h
spectral function is peaked at the AFM wave vector and 
is almost $\omega$-independent for low
frequencies $\omega\ll W_{\text{coh}}$
(\ref{61}), which results in
\begin{equation}
\Gamma(\omega)\simeq
\frac{\pi\Lambda Z^{2}\chi}{8W_{\text{coh}}D\omega_{\text{0}}}
\omega.
\label{90}
\end{equation} 
Unlike the previous case, now the electron 
band structure is  mainly
responsible for obtained behavior.
Of course, the AFM character of spin fluctuations favors the
scatter process with momentum transfer $Q$ and, thus,
enhances its contribution to the relaxation rate.

To summarize the low-energy behavior of optical conductivity,
we have shown that the relaxation rate due to the
electron scattering on spin fluctuations
exhibits the crossover from the $\Gamma(\omega)\sim \omega^{3/2}$
behavior  at low frequencies $\omega \ll 2|\mu|$
to a linear $\omega$-dependence at $\omega > 2|\mu|$.

Now we discuss the conductivity at intermediate frequencies. 
The exact diagonalization studies of the latter quantity have suggested a possible explanation of the MIR absorption
within the one-band model. \cite{daggoto,eder} For instance, as it has been
 observed
in Ref.~\onlinecite{eder}, the finite frequency part of $\sigma(\omega)$ 
is dominated by a single excitation which scales with $J$
in the underdoped regime. \cite{eder} The origin of this excitation was
ascribed to  transitions in which  internal degrees of freedom of the
 spin-bag QP are excited. The presence of  extra absorption ranging from MIR frequency
to $\sim$ 1 eV was also observed in the $1/N$ expansion study of the
$t-J$ model. \cite{kotliar1}
The authors of Ref.~\onlinecite{kotliar1} interpreted this feature as
being  due to the
incoherent motion of charge carriers. Since the existence
of broad incoherent band in the density of states of charge carriers
is due to  internal degrees of freedom of the spin-bag QP
it follows that the underling physics of both 
these points are the same.
Bellow we also support this explanation of the MIR band.

Actually, with increasing energy 
an extra channel of  optical transitions opens.
These are  the transitions which involve an incoherent band of the 
single--particle
spectral function. As we have already discussed, 
the incoherent--coherent transitions are characterized by the energy
scale $\Delta=W_{\text{coh}}-2|\mu|$ being a threshold energy for creating
``particle-hole'' pairs  with a ``hole'' in the incoherent band.
Due to this extra channel at  $\omega>\Delta$, the real part of 
$\sigma(\omega)$ starts to increase. Since  $\sigma(\omega)$ vanishes in the 
limit 
$\omega\rightarrow\infty$,  there should be a peak
in conductivity  at energies of order $\Delta$.
Since the coherent bandwidth $W_{\text{coh}}$ (and hence $\Delta$)
scales with $J$ it follows that 
the typical energy of the peak is also $J$ that coincides with the
energy scale of fine structure in high-energy absorption found
in Ref.\onlinecite{eder}. 

To conclude this section, we discuss the doping dependence of 
the Drude and  regular parts of the optical conductivity. 
The doping dependence of the renormalized Drude weight $\tilde{D}$ (\ref{81})
 is due to the $\delta$--dependence of both kinetic 
energy and mass enhancement factor. 
For small $\delta$ the former one scales as 
$|\!<\!H_{t}\!>\!|\sim ZN_{1}\sim \delta$ and increases with doping. 
While the mass enhancement factor, $1+\lambda$, that is due to the
 electron scattering on spin fluctuations gets smaller upon doping 
since the system moves away from the AFM phase boundary.
Combining both points we conclude that the Drude weight 
increases faster than the number of doped holes.
Since the sum rule for the optical conductivity has to be fulfilled
we expect  transfer of the spectral weight from the regular to the
Drude part of the conductivity.
The loosing of the spectral weight by the mid infrared band is
provided by the fact that the normalized 
rate of the incoherent-coherent transitions 
$\Pi^{\prime\prime}_{1}(\omega)^{\text{i-c}}/D\sim
 Z/D\Gamma\sim (1-\delta)$   decreases      upon doping.
This  qualitative picture of doping dependence of $\sigma(\omega)$
is in general agreement with finite clusters calculations.
To give some quantitative estimates on  evaluation of $\sigma(\omega)$ 
with doping
one needs  a  more detailed information
concerning  $\delta$-dependence of one-particle spectral
characteristics as well as spin fluctuation spectrum  used as the inputs in 
our theory.
However, the presently available numerical data on the $\delta$-dependence of these quantities 
is far from convincing.

\section{Conclusion}

To summarize, we have developed a self-consistent theory  
both for the dynamic charge susceptibility and the optical conductivity
within the memory function formalism in terms of the Hubbard operators.
In this framework the charge susceptibility has been expressed via
the memory function  $M_{\bf q}(\omega)$, that is given by fluctuating force
correlation function. The essential approximation of the preceding theory 
is the mode-coupling approximation. 
Within the MCA the memory function
has been  factorized in terms of single-particle GF and spin and charge susceptibilities.  
As a result a type of the ``golden-rule formula'' has been obtained:
the density fluctuation created at some place and time 
will decay either into p-h pair excitations and/or
p-h pairs and spin (charge) density fluctuations if the system evolves into
the future. Though, at present time the precise range of validity of MCA
can not be given the following general arguments can be used to justify it.
 First, being a self-consistent scheme of calculations
MCA does not use any unphysical for strongly correlated systems the
zero-order GF
and can be  viewed   
as an analog of the non-crossing or the self-consistent Born approximation (SCBA) for the single-particle GF.   
However in the SCBA vertex corrections are neglected and 
only the skeleton loop diagrams are taken into account. 
While MCA correctly reproduces the back-scattering term in the impurity
 problem \cite{gbs} and thus partially accounts for vertex corrections
[see Eqs. (\ref{21}) and  (\ref{22}) for the present case]. 
Next, this approach allows to deal with constrained electron operators
avoiding an auxiliary field representation 
that rises the problem of local constraint.
Moreover, within our decoupling scheme the operators from the same site 
is never decoupled and thus the strong local correlations are retained.
Finally, it is expected  that there is no divergence 
in renormalized vertexes signaling the existence of a critical low energy mode and the vertex renormalizations
that is not accounted by  our theory can change only 
our numerical estimates but not the qualitative behavior of the susceptibility.

Our main findings  are as follows.
We have shown that in the long-wavelength limit the charge fluctuation spectrum
is mainly governed by the sound mode (\ref{55}). Although unrenormalized sound
velocity is larger than  the Fermi velocity, the ``self-energy'' corrections
soften the sound. Sound  falls down into the
the particle-hole continuum and thus acquires a finite damping due to the decay into  pair excitations. The sound damping (\ref{69}) is only
numerically smaller than its energy and hence there is no well--defined sound 
mode. 

At large momenta the density fluctuation spectrum mainly consists of a 
broad high-energy peak which nearly follows the MF dispersion (\ref{38}). 
At wave vectors large enough  the peak is dispersed out of the
 coherent particle-hole continuum  and its broadening (\ref{70}) is 
due to the  high energy $\sim t$  transitions
 involving the incoherent band  of the single--particle excitations.

We have also discussed the optical conductivity. 
At low frequencies we have analyzed $\sigma(\omega)$ in terms of 
the generalized Drude law.
We have shown that there is a large mass enhancement of order 
$m^{*}/m\simeq 6$, 
due to the electron scattering on spin fluctuations. This
scattering process also leads to the non-Drude fall-off of
the low energy part of $\sigma^{\prime}(\omega)$.  
Namely, the relaxation rate shows a power law
$\omega$--dependence  with the exponent $3/2$
at low frequencies  $\omega<2|\mu|$ and it is linear
in $\omega$ at frequencies $\omega>2|\mu|$.
As for the intermediate frequency conductivity, we have pointed out the
existence of a characteristic energy $\Delta$ (of order $J$)
above which an extra channel of the optical transitions opens. These are
the transitions in which  particle-hole pairs with a ``hole'' in the incoherent
band are excited and they might be responsible for the experimentally observed
MIR absorption.

The obtained results are in  good agreement with the exact diagonalization studies of small clusters. \cite{tohoyama}

In this paper we have focused on the moderate doping regime, when
the Fermi surface is large and there exist a strong short-range AFM correlations in the system. Our assumption about the large electronic FS gives 
a lower boundary $\delta_{c}$ of the range of validity of the present
study with respect to the hole doping ($\delta_{c}$ being the hole concentration when the topology of the FS changes, i.e. when  a transition from
large electronic to the hole pocket-like FS takes place when doping becomes smaller than $\delta_{c}$)\cite{FS}. To  discuss the low doping regime 
$\delta<\delta_{c}$ one has to take into account the changes in the QP spectra. Here, we only point out, that for  $\delta<\delta_{c}$
the new low energy scale $v_{\text{F}}\sim J\sqrt{\delta}$
appears in the charge dynamics. Due to this and the different topology of FS one can expect the different behavior of the low energy charge fluctuations that
are driven by FS related particle-hole excitations.
 The existence of the upper boundary is due to the assumption
on the presence of strong short range   AFM correlations.
 When these correlations are already  destroyed 
the assumption that the scattering on spin-fluctuations gives the main input 
to the charge relaxation is not valid anymore  and 
in order to relate the theory
with experiment one has to include also the scattering on other bosonic degrees of freedom. However, this doping regime is beyond the scope of the present
consideration.

\acknowledgments
We would like to thank  V. Kabanov, V. Yushankhai, V. Oudovenko
and N. Perkins for useful discussions and comments.
Financial support by the 
the INTAS--RFBR Program Grant No 95--591 is acknowledged.
One of the authors (N. M.) acknowledges also the support by NREL
in the framework of Subcontract No AAX--6--16763--01.




\begin{references}
\bibitem{irr} The irreducibility means that only the part of ``current''
               operator orthogonal to 
              the density one is retained in the memory function.
               For the rigorous definition see
              Ref.~\onlinecite{tser}.
\bibitem{incom}  Although experiment suggests that the spin susceptibility
                is peaked at the incommensurate wave vectors, we do not 
                discuss this case since it does not introduce qualitatively 
                new physics.
\end{references}
\end{document}